\documentclass[a4paper,11pt,onecolumn,superscriptaddress]{article}

\usepackage{comment}
\usepackage{graphicx}
\usepackage{epsf}
\usepackage{bm}
\usepackage{amsmath}
\usepackage{amsfonts}
\usepackage{amssymb}
\usepackage{epstopdf}
\usepackage{color}
\usepackage[dvipsnames]{xcolor}
\usepackage{verbatim}
\usepackage{multirow}
\usepackage{soul}
\usepackage{bm}

\usepackage[width=0.00cm, height=0.00cm, left=1.50cm, right=1.50cm, top=2.00cm, bottom=2.00cm]{geometry}
\usepackage{microtype}
\usepackage[colorlinks = true,
            linkcolor = teal,
            urlcolor  = black,
            citecolor = teal,
            anchorcolor = blue]{hyperref}
\usepackage[capitalize]{cleveref}
\usepackage[normalem]{ulem}
\usepackage{enumitem}
\usepackage{booktabs}

\usepackage{lipsum}

\makeatletter\let\expandableinput\@@input\makeatother

\begin{document}

\begin{center}
		\vspace{0.4cm} {\large{\bf Investigating the $w$CDM Model with Latest DESI BAO Observations
}} \\
		\vspace{0.4cm}
		\normalsize{ Manish Yadav $^1$, Archana Dixit$^2$, M. S. Barak$^3$,  
         Anirudh Pradhan$^4$ }\\
		\vspace{5mm}
		\normalsize{$^{1,3}$ Department of Mathematics, Indira Gandhi University, Meerpur, Haryana 122502, India}\\
		\normalsize{$^{2}$ Department of Mathematics, Gurugram University, Gurugram- 122003, Harayana, India.}\\ 
        \normalsize{$^{4}$ Centre for Cosmology, Astrophysics and Space Science (CCASS), GLA University, Mathura-281 406, Uttar Pradesh, India}\\
		\vspace{2mm}
		$^1$Email address: manish.math.rs@igu.ac.in\\
		$^2$Email address: archana.ibs.maths@gmail.com\\
            $^3$Email address: ms$_{-}$barak@igu.ac.in\\
            $^4$Email address: pradhan.anirudh@gmail.com\\
\end{center}

{\noindent {\bf Abstract.}
In this study, we explore the impact of various combinations of CMB-independent datasets, including the recent DESI BAO measurements, on the equation of state (EoS) of dark energy and other cosmological parameters within the framework of the dynamical dark energy model ($w$CDM). Assuming a constant EoS parameter for dark energy, we derive constraints on the free parameters of the model using observational datasets such as DESI BAO, BBN, Observational Hubble Data (OHD), and Pantheon Plus (SN$^+$) $\&$ SH0ES. Our analysis examines the deviations of the $w$CDM model from the standard $\Lambda$CDM scenario and assesses its implications for cosmological tensions, particularly the $H_0$ tension [$\text{km} \text{s}^{-1} \text{Mpc}^{-1}$]. We find that the combination of DESI BAO + BBN + OHD + SN$^+$ (DESI BAO + BBN + OHD + SN$^+$ \&SH0ES) datasets provides constraints on $w_{\mathrm{de}0}$, suggesting a possible deviation from the cosmological constant scenario at a significance level of $1.6\sigma$ ($1.4\sigma$), respectively. Furthermore, we observe an inverse correlation between $w_{\text{de0}}$ and $H_0$, which highlights the role of dark energy dynamics in resolving the tension $H_0$ by approximately $2.1\sigma(0.8\sigma)$ from DESI BAO + BBN + OHD (DESI BAO + BBN + OHD + SN$^+$ \&SH0ES)  datasets, respectively.
 Our findings offer valuable insights into the nature of dark energy and its influence on the cosmic expansion history, with implications for future observational efforts. We utilize the Akaike Information Criterion (AIC) and Bayesian Information Criterion (BIC) to evaluate our model's performance. The results indicate that the $w$CDM model demonstrates superior effectiveness.\\

\smallskip 
 {\bf Keywords} :DESI, Obsevational data, $w$CDM model\\
 PACS: 98.80.-k

\section{Introduction}
The accelerated expansion of the universe was first identified in 1998 through observations of the apparent magnitudes of Type Ia supernovae (SNe Ia) \cite{ref1,ref2,ref3}. Subsequent cosmological studies confirmed this phenomenon, particularly through measurements of temperature anisotropies and polarization in the cosmic microwave background (CMB) \cite{ref4,ref5,ref6} radiation, analyses of the large-scale structure (LSS) of the universe \cite{ref7,ref8,ref9,ref10}, assessments of baryon acoustic oscillation (BAO) peak scales \cite{ref11,ref12,ref13,ref13a}, and investigations of the Hubble parameter \cite{ref14,ref15,ref16,ref17}. A widely accepted explanation for this acceleration is the dominance of dark energy or a dark fluid, which possesses an effective negative pressure \cite{ref18,ref19,ref20}. The discovery of dark matter in the universe was first made through the unexpectedly high rotational velocities observed in the outer regions of galaxies \cite{ref21}, making it one of the major unsolved mysteries of modern cosmology. In recent years, there has been a growing agreement among the scientific community that the standard $\Lambda$CDM (Here $\Lambda$ and CDM refer to cosmological constant and cold dark matter associated with dark energy and dark matter) model may be a more realistic cosmological model than other models based on the fundamental principles of inflationary paradigm \cite{ref22, ref23,ref24, ref25}. This model is highly consistent with current constraints on key cosmological parameters, as well as with astronomical observation data sets from the universe \cite{ref26,ref27,ref28,ref29,ref30}. However, in the modern era of high-precision cosmology, several inconsistencies have emerged as statistically significant with various data sets  \cite{ref31,ref32}.\\

The Hubble constant \(H_0\)  is one of the key cosmological parameters, crucial for understanding the Universe's expansion rate, composition, and long-term evolution. However, measurements of \(H_0\) based on observations of the early and late Universe yield conflicting results. Over the past decade, this discrepancy, known as the ``Hubble tension," has emerged as a significant challenge in modern cosmology. Two primary methods for determining \(H_0\) involve SNe Ia and CMB radiation. The SNe Ia approach relies on the astronomical distance ladder in the local Universe, whereas the CMB method derives \(H_0\) from the inverse distance ladder within the framework of the \(\Lambda\)CDM cosmological model. Notably, the Supernova \(H_0\) for the Equation of State (SH0ES) project reported a higher value of \(H_0 = 73.04 \pm 1.04\) km s\(^{-1}\) Mpc\(^{-1}\) \cite{ref33}, while measurements from the Planck Collaboration, based on CMB data, indicated a lower value of \(H_0 = 67.4 \pm 0.5\) km s\(^{-1}\) Mpc\(^{-1}\)  \cite{ref34}. In addition, there are several late-time measurements that suggest a higher value for the Hubble constant, which is in disagreement with the Planck-CMB estimate. For example, the Megamaser Cosmology Project \cite{ref34a} found $H_0 = 73.9 \pm{3.0}  $ kms$^{-1}$Mpc$^{-1}$, while the Surface Brightness Fluctuations \cite{ref34b} gave a value of $H_0 = 73.3 \pm{2.4}  $ kms$^{-1}$Mpc$^{-1}$. However, the CMB data suggest a lower value of $H_0$, which is consistent with the constraints of BAO and BBN, along with the corroborative evidence from other CMB experiments such as ACT-PolDR4 \cite{ref34c}, ACT-PolDR6 \cite{ref34d} and SPT-3G \cite{ref34e}.\\

The discrepancy in $H_0$ measurements, known as the Hubble tension, has now reached a significance level of \(5\sigma-6\sigma\), making it unlikely to be a statistical anomaly. As a result, extensive research is being conducted to determine whether modifications to the standard \(\Lambda\)CDM cosmological model are necessary. Some proposed alternatives include interacting dark energy models \cite{ref35,ref36,ref37,ref37a,ref37b} and f(T) gravity theories \cite{ref38,ref39}. In parallel, significant efforts have been devoted to addressing potential systematic errors in astrophysical distance measurements that could influence the determinations \(H_0\) discussed in \cite{ref40,ref41}. Developing cosmological model-independent methods for estimating \(H_0\) has become particularly important. For example, J. L. Bernal et al. (2016) \cite{ref42} reconstructed the late-time expansion history using BAO and SN Ia data to infer \(H_0\). However, their approach depends on the sound horizon scale \(r_d\) at the radiation drag epoch, introducing an additional parameter into the analysis. More recently,  Renzi and Silvestri \cite{ref43} proposed a method to constrain \(H_0\) based solely on the distance duality relation (DDR) alongside direct observational data, including SN Ia, BAO, and Observational Hubble Data (OHD). Motivated by the latest data release from the Dark Energy Spectroscopic Instrument (DESI) survey, this study builds on their methodology to update \(H_0\) constraints at different redshifts, aiming to uncover new insights into resolving the Hubble tension. Furthermore, it's important to note that the $H_0$ tension is actually a tension on the absolute magnitude $M_B$ of Type Ia supernovae (SNe Ia) because the SH0ES $H_0$ measurement comes directly from $M_B$ estimates. These outcomes have provided valuable opportunities to modify the view on the design of the standard cosmological model, allowing for exploration beyond the $\Lambda$-Cold Dark Matter framework. Thus, even though a wide number of dark energy models have been proposed to address the $H_0$ tension and $M_B$ tension such as sign- switch dark energy model based on graduated dark energy, dark matter, and dark energy interaction. This concordance paradigm has consistently passed all experimental tests and appeared to be statistically favored.\\

The BAO Collaboration has released the first yearly observation data of the DESI, which revealed a $2\sigma$ cosmological discrepancy with the dynamical dark energy model upon initial examination \cite{ref44}. The $w_0w_a$CDM model has also provided the same evidence for dynamical dark energy with different analyses using various types of datasets, as seen in ref \cite{ref45,ref46}. In addition to these findings, the DESI BAO  data, combined with other astrophysics data sets such as CMB, CC, and SNIa observations, have been used to constrain cosmological parameters in various cosmological models \cite{ref47,ref48,ref49,ref50,ref51,ref52,ref53,ref54,ref55,ref56}. However, the DESI BAO collaboration found that the equation of state (EoS) of dark energy in the $w$CDM model slightly favors the quintessence region $(w_{de0}>-1)$ with DESI+CMB+DESY5 data, while also slightly favoring the phantom region $(w_{de0}<-1)$ with DESI+CMB data \cite{ref57}, we notice that the EoS of dark energy strongly favors cosmological constant with CMB and other combination of datasets. Due to this, there is no evidence of dynamical dark energy in $w$CDM model. We use a CMB-independent approach to constrain cosmological parameters and also discuss the deviation of EoS of dark energy. \\ 

The DESI is carrying out a Stage IV survey \cite{ref58, ref59} aimed at refining cosmological constraints by studying the clustering of galaxies, quasars, and the Lyman-$\alpha$ forest. Spanning a five-year period, DESI is mapping 14,200 square degrees across a redshift range of 0.1 to 4.2, utilizing a spectroscopic sample significantly larger than previous SDSS surveys. The survey categorizes six different tracers, including low-redshift galaxies from the Bright Galaxy Survey (BGS), luminous red galaxies (LRG), emission line galaxies (ELG), quasars as direct tracers, and Lyman-$\alpha$ forest quasars for mapping neutral hydrogen distribution. Additionally, it includes a high-density stellar target sample as part of an overlapping Milky Way Survey \cite{ref60} to study stellar evolution and galactic dynamics. For cosmological analysis, DESI is designed to provide precise measurements of the universe's expansion history and large-scale structure formation. Early observations have confirmed the BAO \cite{ref61} signal with a few percent accuracy, demonstrating that the survey is progressing as planned. More specifically, DESI \cite{ref62} will deliver stringent constraints on fundamental cosmological parameters, including matter density, the dark energy EoS, spatial curvature, the amplitude of primordial fluctuations, and neutrino mass. Moreover, it will rigorously examine potential modifications to general relativity proposed to explain the accelerated expansion of the universe \cite{ref63, ref64, ref65}.

In this paper, we have assumed a constant EoS parameter of dark energy. Our aim in this study is to investigate how various combinations of CMB independent data sets, including the recent  DESI BAO, affect EoS of dark energy and other cosmological parameters with a dynamical dark energy model ($w$CDM). We  derive the constraints on the free parameters of the model under consideration. The structure of the paper can be summarized as follows: In  section I, we have  present the  complete overview of the model. In section-II we have discussed the observational dataset, analysis and methodology. In section III, we present the results and discuss the findings of this analysis. Finally, we conclude with a summary of our work in section IV.

\section{MODEL, DATA AND METHODOLOGY}\label{sec2}


\label{sec:datasets}

 One of the promising dynamical dark energy models proposed, known as the $w$CDM model. As an extension of the $\Lambda$CDM model, it replaces $\Lambda$ with a scalar field characterized by a constant barotropic factor, $w$. The governing  Friedmann equation for $w$CDM model reads as:

\begin{equation}
    \frac{H^2(z)}{H_0^2}  =\Omega_{\text{r}0}[1+z]^{4}+\Omega_{\text{m}0}[1+z]^{3}+\Omega_{\rm de0}[1+z]^{3(1+w_{\rm de0})},
\end{equation}
Here, $\Omega_{r0}$, $\Omega_{m0}$,  and $\Omega_{de0}$ denote the present density parameters of radiation, matter,  and dark energy, respectively. These parameters satisfy the equation 
   $ \Omega_{r0}+ \Omega_{m0}+  \Omega_{de0} = 1$.\\

   The datasets and methodology used are as follows:
   
   \begin{itemize}

\item \textbf{DESI BAO}:  

The recently published BAO measurements from the DESI collaboration incorporate data from six different tracers: quasars (QSO), Lyman-$\alpha$ forest quasars (Lya QSO), the bright galaxy survey (BGS), luminous red galaxies (LRG), emission line galaxies (ELG), and a combined LRG+ELG sample. These trasers can provide precise constraints on three important quantities \cite{ref66,ref67,ref44}, namely the Hubble horizon $D_H(z)/r_{\rm d}$, the transverse comoving distance $D_{M}(z)/r_{\rm d}$ and the
angle-averaged distance $D_{v}(z)/r_{\rm d}$  are presented in Table \ref{tab1} , where $r_{\rm d}=\int_{z_{\rm d}}^\infty \frac{c_{\rm s}\text{d}z}{H(z)}$ is sound horizon at the drag redshift ($z_{\rm d}$) and ($c_{\rm s}$) is  sound speed  of the baryon–photon fluid.\\

The transverse comoving distance for flat ($\kappa = 0$) universe is defined  as,

  \begin{equation}
 	D_{M}(z)=  \int_0^z \text{d}z' {c \over H(z')}.\\
 \end{equation}

and the Hubble distance is caluclated as,

\begin{equation}
     D_H(z) = \frac{c}{H(z)},
\end{equation}

 The angle-averaged distance is defined as
 
 \begin{equation}
  D_V(z) \equiv \left[z D^2_M(z) D_H(z)\right]^{1/3}
 \end{equation}

We quantify the goodness-of-fit of our model by defining a statistical $\chi^2$ function based on the  analysis of DESI BAO mearurement read as,

\begin{equation}
	\chi^2_{\text{DESI BAO}} =  \Delta Q_i \left(C_{\text{DESI BAO}}^{-1}\right) \Delta {Q_i}^T.
\end{equation}

where,
\[
\Delta Q_i =
\begin{cases}
	\left( \frac{D_M}{r_d} \right)^{\mathrm{th}}(z_i) - \left( \frac{D_M}{r_d} \right)^{\mathrm{obs}}(z_i), & \text{for } D_M \text{ measurements}, \\[8pt]
	\left( \frac{D_H}{r_d} \right)^{\mathrm{th}}(z_i) - \left( \frac{D_H}{r_d} \right)^{\mathrm{obs}}(z_i), & \text{for } D_H \text{ measurements}, \\[8pt]
	\left( \frac{D_V}{r_d} \right)^{\mathrm{th}}(z_i) - \left( \frac{D_V}{r_d} \right)^{\mathrm{obs}}(z_i), & \text{for } D_V \text{ measurements}.
\end{cases}
\]

and $C_{\text{DESI BAO}}^{-1}$ is the inverse of the covariance matrix corresponding to the DESI BAO dataset, which accounts for statistical correlations between BAO measurements.

\begin{table}[ht!]

\caption{\rm The statistics measurement of DESI BAO samples used in ref \cite{ref44}}.
\centering
\begin{tabular}{l|c|c|c|c}

\hline
\textbf{tracer} & $\bm{\,\,z_{\rm eff}\,\,}$  &  $\bm{\,\,D_{\rm V}(z)/r_{\rm d}\,\,}$ & $\bm{\,\,D_{\rm M}(z)/r_{\rm d}\,\,}$ & $\bm{\,\,D_{\rm H}(z)/r_{\rm d}\,\,}$  \\
\hline

\hline
``BGS & $0.30$ & $7.93 \pm 0.15$ & --- & --- \\

LRG & $0.51$ & --- & $13.62 \pm 0.25$ & $20.98 \pm 0.61$ \\

LRG & $0.71$ & --- & $16.85 \pm 0.32$ & $20.08 \pm 0.60$ \\

 LRG + ELG & $0.93$ & --- & $21.71 \pm 00.28$ & $17.88 \pm 0.35$ \\

ELG & $1.32$ & --- & $27.79\pm0.69$ &$13.82\pm00.42$\\

QSO & $1.49$ & $26.07\pm0.67$ & -- & --- \\

Lya QSO & $2.33$ & --- &$39.71\pm0.94$ & $8.52 \pm 0.17$" \\
\hline
\hline
\end{tabular} 
\label{tab1}
\end{table}

\item\textbf{Observatinal Hubble Data (OHD)}: We explore 33 measurements of $H(z)$ from OHD at $ z \in (0.07,1.965) $, as presented in Table \ref{tab2}.  The foundational idea, originating from \cite{ref68}, establishes the Hubble parameter $H(z)$ in terms of  redshift $z$, and cosmic time $t$ read 
	
\begin{equation}
H(z)= -{(1+z)}^{-1}\frac{\text{d}z}{\text{d}t}.
\end{equation}
We will define the chi-squared function, signified by $\chi^2_{\rm OHD}$, for these measurements as follows:
\begin{equation}
\chi^2_{\rm OHD} = \sum_{i=1}^{33} \frac{[H^{\text{obs}}(z_i)-H^{\text{th}}(z_i)]^2}{\sigma^2_{H^{\text{obs}}(z_i)}},
\end{equation}
where $H^{\text{obs}}(z_i)$ and $H^\text{th}(z_i)$ stands for the observed  and theoretical value of  Hubble parameter  at redshift $z_i$, while the standard deviation $\sigma^2_{H^{\text{obs}}(z_i)}$  provided in the aforementioned Table \ref{tab2}.\\

\begin{table}[t!]
\caption{\rm Compilation of OHD measurements of $H(z)$.}

\begin{center}
\begin{tabular}{lllr}
\multicolumn{4}{c}{{}}\\
\hline \toprule\hline
$z$ & $H(z)$ & $\sigma_{H(z)}$ & Ref.\\
\hline

0.070 & 69.00 & 19.60 &  \cite{ref69}\\
0.090 & 69.00 & 12.00 & \cite{ref70}\\
0.120 & 68.60 & 26.20 & \cite{ref69}\\
0.170 & 83.00 & 8.00 & \cite{ref70}\\
0.179 & 75.00 & 4.00 &  \cite{ref71}\\
0.199 & 75.00 & 5.00 &  \cite{ref71}\\
0.200 & 72.90 & 29.60 &  \cite{ref69}\\
0.270 & 77.00 & 14.00 & \cite{ref70}\\
0.280 & 88.80 & 36.60 &  \cite{ref69}\\
0.352 & 83.00 & 14.00 & \cite{ref71}\\
0.380 & 83.00 & 13.50 &  \cite{ref72}\\
0.400 & 95.00 & 17.00 & \cite{ref70}\\
0.4004 & 77.00 & 10.20 & \cite{ref72}\\
0.425 & 87.10 & 11.20 &  \cite{ref72}\\
0.445 & 92.80 & 12.90 &  \cite{ref72}\\
0.470 & 89.00 & 49.60 &  \cite{ref73}\\
0.4783 & 80.90 & 9.00 &  \cite{ref70}\\
0.480 & 97.00 & 62.00 & \cite{ref15}\\
0.593 & 104.00 & 13.01 &  \cite{ref71}\\
0.680 & 92.00 & 8.00 &  \cite{ref71}\\
0.750 & 98.80 & 33.60 &  \cite{ref74}\\
0.781 & 105.00 & 12.00 &  \cite{ref71}\\
0.800 & 113.10 &  15.10  &  \cite{ref75}     \\
0.875 & 125.00 & 17.00 &  \cite{ref71}\\
0.880 & 90.00 & 40.00 &  \cite{ref15}\\
0.900 &  117.00 &  23.00 &  \cite{ref70}\\
1.037 & 154.00 & 20.01 &  \cite{ref71}\\
1.300 & 168.00 & 17.00 &  \cite{ref70}\\
1.363 & 160.00 & 33.60 &  \cite{ref76}\\
1.430 & 177.00 & 18.00 &  \cite{ref70}\\
1.530 & 140.00 & 14.00 &  \cite{ref70}\\
1.750 & 202.00 & 40.00 &  \cite{ref70}\\
1.965 & 186.50 & 50.40 & \cite{ref76}\\
\hline 
\bottomrule\hline
\end{tabular}
\label{tab2}

\end{center}
\end{table}

\item \textbf{Big Bang Nucleosynthesis (BBN)}
An alternative way to estimate the density of baryons is by BBN,
which helps us understand the dynamics of the early universe. By using BBN, we can identify the limitations of the conventional cosmological models. We have conducted a new evaluation of the physical baryon density, which is represented as $\omega_b$ (where $\omega_b \equiv \Omega_bh^2$)), and obtained a value of $0.02233\pm0.00036$ derived from BBN. For this calculation, we have incorporated the latest experimental nuclear physics data obtained from the LUNA laboratory in Italy \cite{ref77}.

\item \textbf{PantheonPlus \& SH0ES (SN$^{+}$\& SH0ES)}:
We integrate the latest SH0ES Cepheid host distance calibrations \cite{ref78} into the likelihood function by incorporating distance modulus data from Type Ia supernovae (SNe Ia) within the Pantheon+ compilation \cite{ref79}. The Pantheon (+) dataset consists of 1701 light curves corresponding to 1550 distinct SNe Ia events, spanning a redshift range of \( z \in [0.001, 2.26] \).\\

Our baseline parameters of dynamical dark energy model ($w$CDM) are given by $\mathcal{P}_{w\text{CDM}}
 = \{\omega_b, \omega_{cdm}, H_0, w_{deo}\}$. Where, the baryon energy density $\omega_{\rm b} = \Omega_{\rm b}h^2$, the cold dark matter energy density $\omega_{\rm cdm} = \Omega_{\rm cdm}h^2$, the Hubble constant $H_0$ and the EoS of dark energy $w_{de0}$. We use flat priors for all parameters in our statistical analyses: $\omega_{\rm b}\in[0.018,0.024]$, $\omega_{\rm c}\in[0.10,0.14]$, $H_0\in[60,80]$, and $ w_{de0}\in[-3,1]$. We apply Markov Chain Monte Carlo (MCMC) analyses to constrain the model parameters, using a tailored version of the CLASS+MontePython software \cite{ref80,ref81}. We ensured the MCMC chains converged using the Gelman-Rubin criterion ${R-1 < 0.01}$ and analyzed the results with the GetDist Python module \cite{ref82,ref83}.

\end{itemize}

\section{Results and discussion}
The Fig.\ref{fig1} presents posterior distribution functions (PDFs) of the $H_{0}$, derived from different cosmological probes. The figure shows multiple probability distributions for $H_{0}$ corresponding to different redshifts ($z$) and observational datasets. The peak of each distribution represents the most likely value of $H_{0}$ inferred from that particular dataset. The overlap and non-overlapping regions of the probability distributions highlight how different methods provide independent but sometimes conflicting measurements. The vertical dashed lines labeled Planck, TRGB, and SH0ES represent key $H_{0}$ measurements. Planck (CMB-based) typically predicts a lower $H_{0}$, while SH0ES (Cepheid-based) gives a higher, highlighting the Hubble tension. The overlap (or lack thereof) between different distributions illustrates the degree of agreement between different methods.\\

\begin{figure}
    \centering
    \includegraphics[width=0.7\linewidth]{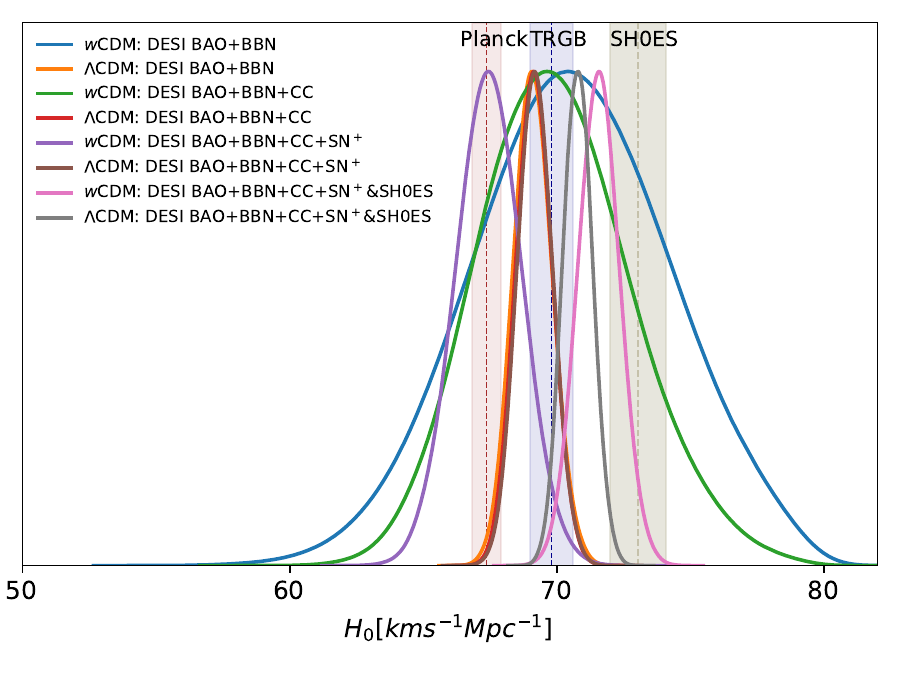}
    \caption{The posterior distribution functions (PDFs) of the Hubble constant ($H_{0}$) are presented, with shaded regions indicating the constraints on $H_{0}$ along with their corresponding 1$\sigma$ intervals. These constraints are derived from SH0ES (Brout et al. 2022), SNa Ia, BBN, and DESI 2024.}
    \label{fig1}
\end{figure}

In Table \ref{tab3}, we investigate the constraints on the baseline free parameters \((\omega_b, \omega_{cdm}, H_0, w_{de0})\) and specific derived parameters  obtained from $w$CDM and $\Lambda$CDM models, based on data combinations such as DESI BAO+BBN, DESI BAO+BBN+OHD, DESI BAO+BBN+OHD+SN$^{+}$, and DESI BAO+BBN+OHD+SN$^+$\&SH0ES. At first, we assessed the impact of the EoS of dark energy with $w$CDM model from all considered datasets. We observe from Table \ref{tab3} that the EoS of dark energy (having $w_{de0}=-0.932 \pm 0.04$ at 68\% and $w_{de0}=-0.932^{+0.07}_{-0.08}$ at 95\%) does not favour of cosmological constant foam of dark energy from DESI BAO+BBN+OHD+SN$^+$. Again, for this model, at 68\% C.L., we have \( w_{\text{de0}}= -1.060^{+0.14}_{-0.14}(-1.030^{+0.12}_{-0.09}) \) with DESI BAO+BBN (DESI BAO+BBN+OHD) data sets, with the universe behaving like a phantom form of the dark energy. At the 95\% confidence level, using the DESI BAO + BBN + OHD + SN$^+$ dataset gives \( w_{\mathrm{de}0} = -0.932^{+0.07}_{-0.08} \), which indicates a quintessence regime. However, when the SH0ES data is added to this combination, the value shifts to 
 	\( w_{\mathrm{de}0} = -1.047^{+0.06}_{-0.06} \), pointing towards a phantom regime. Overall, we conclude that the combination of DESI BAO data with other datasets does not provide strong evidence for an EoS of dark energy parameter deviating significantly from $-1$. Meanwhile, $1.6\sigma$ deviating of $w_{de0}$ from cosmological constant with DESI BAO+BBN+OHD+SN$^+$ datasets. We also observed that $w_{de0}$ inversely correlation with $H_0$ in Fig \ref{fig2}; this means a decreased value of $w_{de0}$ would demand increase value of $H_0$.

\begin{figure*}[hbt!]
    \centering
    \includegraphics[width=0.75\linewidth]{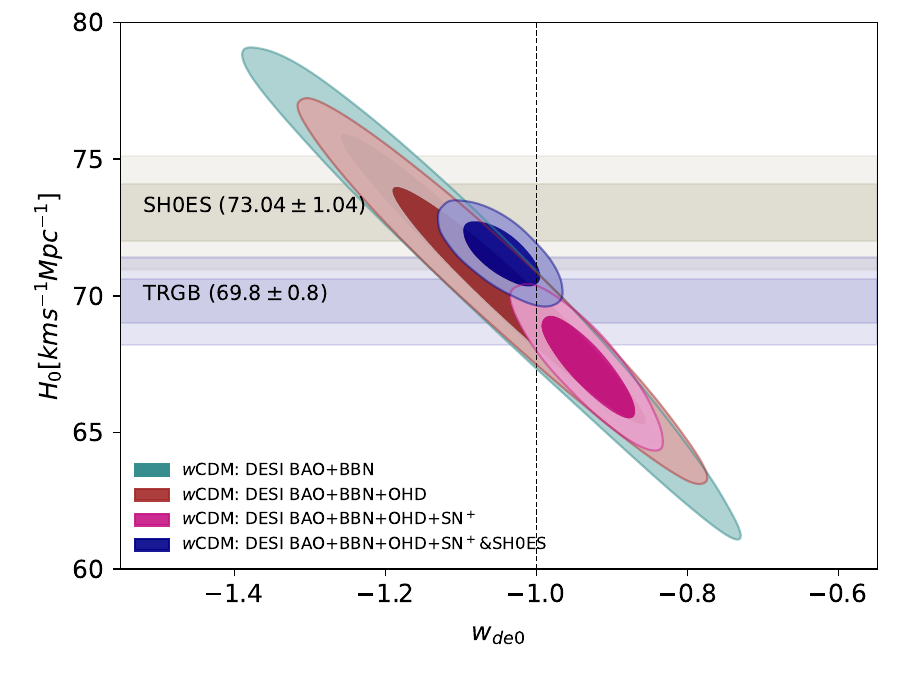}
    \caption{2-dimensional marginalized confidence regions at 68\% and 95\% C.L of $H_0$ and $w_{de0}$ for the $w$CDM 
models from all combination data sets are presented in Table \ref{tab3}. The vertical dotted black line represents $w_{de0} = -1$.}
    \label{fig2}
\end{figure*}

In the modern era of cosmology, one of the most compelling tensions is an $H_0$ tension between Planck data under the $\Lambda$CDM model and direct local measurements from SH0ES, as discussed in Section I. Now, we investigate the $H_0$ constraint results derived by $w$CDM and $\Lambda$CDM models with all given datasets. From Table \ref{tab3}, we notice that the $w$CDM model consistently predicts a higher mean value of   $H_0$  compared to the  $\Lambda$CDM model from DESI BAO+BBN and  DESI BAO+BBN+OHD.  The constraint on $H_0$ reads as: $70.50 \pm 3.70(69.10 \pm 0.74 )$ $\text{km} \text{s}^{-1} \text{Mpc}^{-1}$ under $w$CDM($\Lambda$CDM), from DESI BAO+BBN, respectively, and it reads as: $70.00^{+2.60}_ {-3.10 }(69.15 \pm 0.69)$ $\text{km} \text{s}^{-1} \text{Mpc}^{-1}$  in $w$CDM($\Lambda$CDM) from DESI BAO+BBN+OHD, respectively. Interesting results were found when SN$^+$ data are added to DESI BAO+BBN+OHD datasets; the $w$CDM model has a lower mean value $H_0$ compared to $\Lambda$CDM while the $w$CDM model has a greater mean value $H_0$ compared to $\Lambda$CDM when SH0ES data combine with DESI BAO+BBN+OHD+SN$^+$. The constraint on $H_0$ reads as: $ H_0 = 67.40 \pm 1.20(69.18 \pm 0.67 )$ $\text{km} \text{s}^{-1} \text{Mpc}^{-1}$ under $w$CDM($\Lambda$CDM), from DESI BAO+BBN+OHD+SN$^+$, respectively, and it reads as: $H_0 = 71.56 \pm 0.79(70.78 \pm 0.56)$ $\text{km} \text{s}^{-1} \text{Mpc}^{-1}$  in $w$CDM($\Lambda$CDM) from DESI BAO+BBN+OHD+SN$^+$\&SHOES, respectively. Quantifying the $H_0$ tension with SH0ES data $(H_{0}^{R22}=73.04\pm1.04$ $ \text{km} \text{s}^{-1} \text{Mpc}^{-1})$, we compare the results from the $w$CDM and $\Lambda$CDM models using the DESI BAO+BBN+OHD dataset. The $\Lambda$CDM model shows a significant $H_0$ tension of approximately $3.1\sigma$, while the $w$CDM model exhibits a $H_0$ tension of about $1\sigma$. This indicates that the $w$CDM framework alleviates the $H_0$ tension by approximately $2.1\sigma$ relative to $\Lambda$CDM. With the inclusion of DESI BAO+BBN+OHD+SN$^+$\&SHOES data, a $1.1\sigma$ tension is observed
in the $w$CDM model, compared to $1.9\sigma$ for $\Lambda$CDM; in contrast, the tension is reduced by approximately $0.8\sigma$.

\begin{table*}[hbt!]
     \caption{ The marginalized constraints, presented as mean values with 68$\%$ confidence levels (CL), on both the free and select derived parameters of the $w$CDM and $\Lambda$CDM models for various datasets combinations.}
     \label{tab3}
     \scalebox{0.85}{
 \begin{tabular}{lcccc}
  	\hline
    \toprule
   \textbf{Dataset }&\;\textbf{DESI BAO+BBN}\;&\; \textbf{DESI BAO+BBN+OHD}\;& \;\;\;\textbf{DESI BAO+BBN+OHD+SN$^+$}\; & \textbf{DESI BAO+BBN+OHD+SN$^+$\&SHOES}     
   \\ \hline
      \textbf{Model} & \textbf{$\bm{w}$CDM}\,&\textbf{$\bm{w}$CDM}\,&\textbf{$\bm{w}$CDM}\,&\textbf{$\bm{w}$CDM}\vspace{0.1cm}\\
&\textcolor{teal}{\textbf{$\bm{\Lambda}$CDM}}\, & \textcolor{teal}{\textbf{$\bm{\Lambda}$CDM}}\, & \textcolor{teal}{\textbf{$\bm{\Lambda}$CDM}}\, & \textcolor{teal}{\textbf{$\bm{\Lambda}$CDM}} 
          \\ \hline

\vspace{0.1cm}
{\boldmath$10^{2}\omega_{b}$}&{$2.234\pm 0.036$ }&$2.233 \pm 0.035 $&$2.236 \pm 0.035 $ &$2.252\pm 0.036$\\
 
&\textcolor{teal}{$2.234\pm 0.036$} &\textcolor{teal}{$2.234\pm 0.036$} &\textcolor{teal}{$2.232\pm 0.035$} &\textcolor{teal}{$2.261\pm 0.035$}\\

\vspace{0.1cm}
{\boldmath$\omega{}_{\rm cdm }$}&$0.1210^{+0.0150}_{-0.0140}   $ &$0.1204\pm 0.0098$ &$0.1142\pm 0.0086          $& $0.1378\pm 0.0071$\\

&\textcolor{teal}{$0.1163^{+0.0072}_{-0.0085}$} &\textcolor{teal}{$0.1189\pm 0.0071$} & \textcolor{teal}{$0.1248\pm 0.0059$} &\textcolor{teal}{$0.1327\pm 0.0059$} \\

\vspace{0.1cm}
{\boldmath$H_0$ [$\text{km} \text{s}^{-1} \text{Mpc}^{-1}$]}&$70.50\pm 3.70               $&$ 70.00^{+2.60}_{-3.10} $& $67.40\pm 1.20               $&$71.56\pm 0.79$ \\

& \textcolor{teal}{$69.10\pm 0.74$}&\textcolor{teal}{$69.15\pm 0.69$}  & \textcolor{teal}{$69.18\pm 0.67$} &\textcolor{teal}{$70.78\pm 0.56 $}\\

\vspace{0.1cm}
{\boldmath$w_{de0}$} &$-1.060\pm 0.140             $ &$-1.030^{+0.120}_{-0.099}$ &$-0.932\pm 0.041           $ & $-1.047\pm 0.034$\\

 &-& \textcolor{teal}{-}& \textcolor{teal}{-} &\textcolor{teal}{ -}  \\

 \vspace{0.1cm}
{\boldmath$M_B{\rm[mag]}$}&-  &- &$-19.431\pm 0.036$ & $-19.318\pm 0.021$\\

&- &- & \textcolor{teal}{$-19.385\pm 0.022$} &\textcolor{teal}{$-19.335\pm 0.017 $}  \\

\vspace{0.1cm}
{\boldmath$\Omega{}_{m }  $}&$0.287\pm 0.015            $ &{$0.292\pm 0.017$}  &$0.300\pm 0.012            $& $0.313\pm 0.011$ \\
  
&\textcolor{teal}{$0.290^{+0.013}_{-0.015}   $}& \textcolor{teal}{$0.295\pm 0.014$} & \textcolor{teal}{$0.307\pm 0.011            $} &\textcolor{teal}{$0.310\pm 0.011$} \\

\hline

\vspace{0.1cm}
{\boldmath$\rm{ln} \mathcal{Z}$}&$ -6.88$&$-14.08$&$-728.22 $ &$-675.69$ \\
  
& \textcolor{teal}{$-6.78$}& \textcolor{teal}{$-14.05$} &\textcolor{teal}{$-727.36$}&\textcolor{teal}{$-674.18$}\\

\vspace{0.1cm}
{\boldmath${\rm ln} \mathcal{B}_{ij}$}  &$0.10$ & $0.03$&$0.86$ &$1.69$\\
 \hline
 \hline
\end{tabular}
}
\end{table*}

\color{black}

\begin{figure*}[hbt!]
    \centering
    \includegraphics[width=1.02\linewidth]{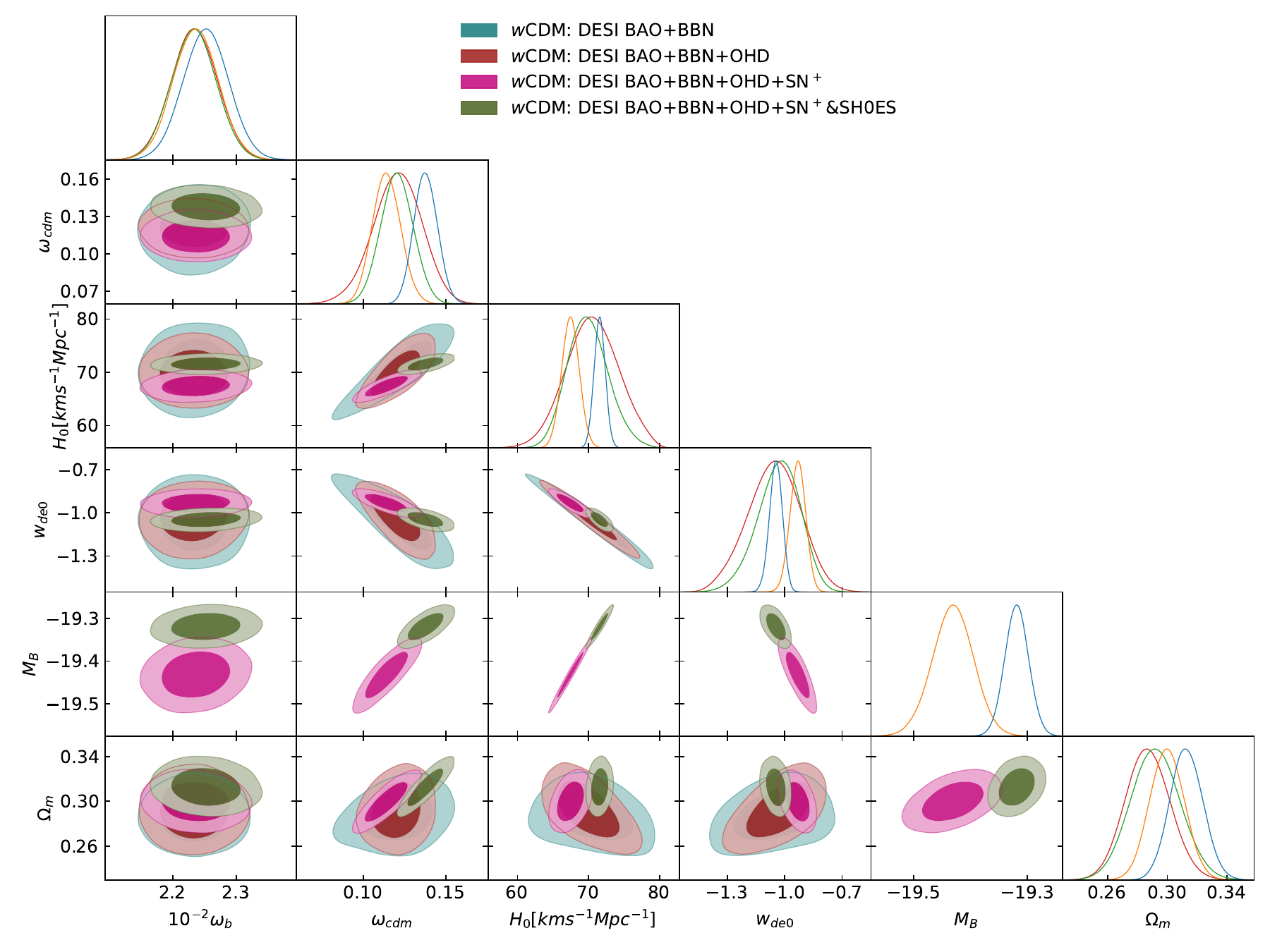}
    \caption{One-dimensional and two-dimensional marginalized contour plots at 68\% and 95\% C.L for various parameters obtained from
 DESI BAO+BBN, DESI BAO+BBN+OHD, DESI BAO+BBN+OHD+SN$^+$, and DESI BAO+BBN+CC+SN$^+$\&SH0ES datasets with the $w$CDM model.}
    \label{fig3}
\end{figure*}
From Fig.\ref{fig3}, we observe a slight negative correlation between $w_{de0}$ and $M_B$ in both models with DESI BAO+BBN+OHD +SN$^+$ and  DESI BAO+BBN+OHD+SN$^+$\&SH0ES datasets. This indicates that higher values of $w_{de0}$ are associated with lower $M_B$ values. In the $w$cdm, the $M_B$ values are constrained as $M_B = -19.431 \pm 0.036(-19.431 \pm 0.036)$ from DESI BAO+BBN+OHD+SN$^+$  (DESI BAO+BBN+OHD+SN$^+$\&SH0ES) datasets, respectively. In this contrast, we notice that $M_B$  tension alleviates in $w$CDM compared to the $\Lambda$CDM model approximately $0.5\sigma$, from DESI BAO+BBN+CC+SN$^+$\&SH0ES. Finally, we assess which model is more effective by calculating the Bayesian evidence and applying the Jeffreys' scale for model comparison. The last two rows of Table \ref{tab3} display the log-Bayesian evidence ($\ln \mathcal Z$) value for $w$CDM and $\Lambda$CDM models, as well as the Bayes’ factor  ($\ln \mathcal{B}_{ij} =| \ln \mathcal{Z}_{\Lambda \rm CDM} - \ln \mathcal{Z}_{w\rm CDM}|$), which quantifies the difference in log-Bayesian evidence between $\Lambda$CDM model and the $w$CDM model. We categorize the strength of the evidence as follows: it is considered inconclusive when $0 \leq | \ln B_{ij}|  < 1$, weak if $1 \leq | \ln B_{ij}|  < 2.5$, moderate if $2.5 \leq | \ln B_{ij}|  < 5$, strong if $5 \leq | \ln B_{ij}|  < 10$, and very strong if $| \ln B_{ij} | \geq 10$. Based on Bayesian evidence, Table \ref{tab3} indicates an inconclusive comparison between the $w$CDM and $\Lambda$CDM models across  DESI BAO+BBN, DSI BAO+BBN+OHD, and DESI BAO+BBN+CC+SN$^+$ datasets. Also, we find weak Bayesian evidence between $w$CDM and $\Lambda$CDM models from  DESI BAO+BBN+CC+SN$^+$\&SH0ES  datasets.\\

Fig.\ref{fig4} presents the marginalized posterior constraints on the matter density parameter ($\Omega_{m}$) and the Hubble constant ($H_{0}$), derived from the combination of BAO data with external datasets used to calibrate the BAO characteristic scale. The figure illustrates the constraints on $\Omega_{m}$ and $w$ obtained from the latest DESI and DESI BAO measurements and their combinations. Using BAO data alone, the matter density parameter is found to be $\Omega_{m} = 0.274 \pm 0.020$, which aligns with the DESI result of $\Omega_{m} = 0.295 \pm 0.015$. However, incorporating the full dataset yields $\Omega_{m} = 0.313 \pm 0.007$, which is in excellent agreement with the values reported by Planck.
\begin{center}
    \begin{figure*}[hbt!]
    \centering
    \includegraphics[width=0.7\linewidth]{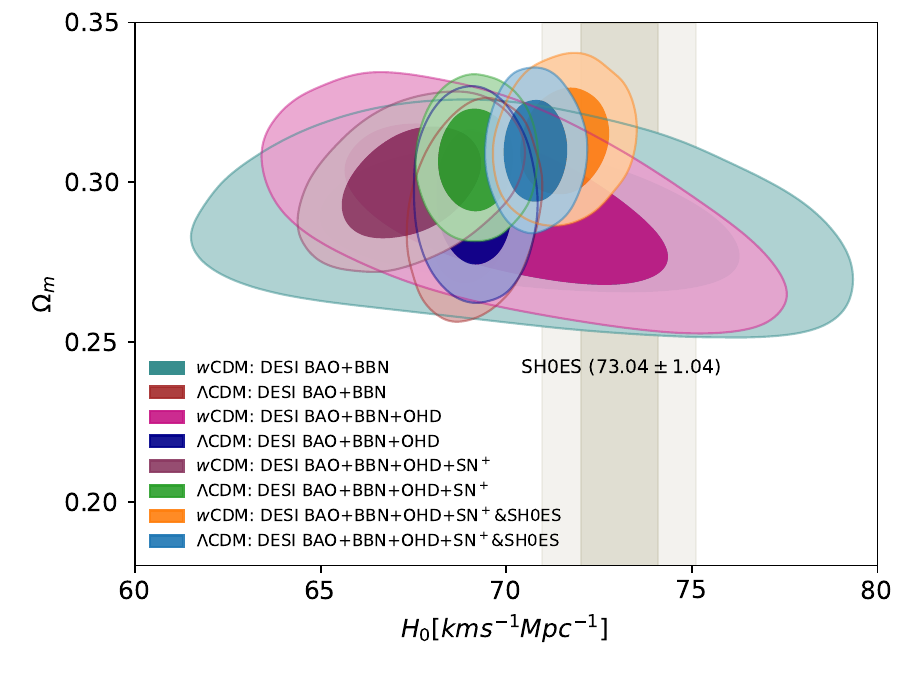}
    \caption{Two-dimensional marginalized confidence regions at 68\% and 95\% C.L of $\Omega_m$ and $H_0$ for the $w$CDM and $\Lambda$CDM 
models from all considered combination data sets are presented in Table \ref{tab3}. The vertical light brown band  represents SH0ES $(H_0=73.04 \pm 1.04)$ measurement.}
    \label{fig4}
\end{figure*}
\end{center}

In this section, we also  incorporated non-BAO datasets specifically OHD, Pantheon+ SNe Ia, and SH0ES calibrations in addition to DESI BAO and BBN, to assess their combined impact on the $w_{\rm de0}$ parameter and the $H_0$ tension. The inclusion of OHD data provides direct $H(z)$ constraints at intermediate redshifts, which slightly shifts the preferred $w_{\rm de0}$ towards the phantom regime ($w_{\rm de0} < -1$) in some combinations, while SN$^+$ data tends to pull it closer to the quintessence regime ($w_{\rm de0} > -1$). Adding SH0ES data further favours a phantom-like EoS.Importantly, these combinations alter the inferred $H_0$ values and thus the level of $H_0$ tension: with DESI BAO+BBN+OHD, the $w$CDM model reduces the tension with SH0ES from $\sim3.1\sigma$ (in $\Lambda$CDM) to about $1\sigma$, a $\sim2.1\sigma$ alleviation. Even with SN$^+$ included, the $w$CDM framework produces mean $H_0$ values that differ systematically from $\Lambda$CDM, demonstrating a sensitivity to low-$z$ distance ladder information.Thus, the inclusion of non-BAO datasets produces measurable shifts in both $w_{\mathrm{de0}}$ and the $H_0$ tension, demonstrating how multiple observational windows jointly constrain dark energy dynamics and the late-time expansion history.These results, although statistically moderate, demonstrate that the inclusion of non-BAO datasets not only influences the central estimates of $w_{\rm de0}$ but also has a measurable impact on the $H_0$ tension. This multi-probe approach thus provides a clearer picture of how different observational data constrain dark energy dynamics and cosmic expansion history.

\begin{center}
    \textbf{{STATISTICAL MODEL COMPARISON }}

\end{center}

To determine the most suitable model, we employ two widely recognized criteria in model selection: the Akaike Information Criterion (AIC) and the Bayesian Information Criterion (BIC). These criteria are extensively utilized in statistics and data analysis, each applying distinct penalties for the inclusion of additional parameters, which may result in different model selection outcomes. The AIC is a technique used to identify the best-fitting model by considering both the likelihood function of the data and the complexity of the model. It penalizes models with a larger number of parameters to mitigate the risk of overfitting while prioritizing models that provide a better fit to the data. 
We fit the data to optimize the model parameters, and then examine at which of those values are supported by the observations.To accomplish so, we employ standard information criteria, specifically the AIC \cite{ref84} and the BIC  \cite{ref85}. These criteria not only consider the value of $\chi^{2}$ and min, but also the number of free parameters. This is especially important when comparing models without nesting. However, BIC is more stringent than AIC, especially for larger datasets, as it penalizes model complexity more heavily. In particular, a model with lower $\triangle$AIC or $\triangle$BIC indicates that it is more favored by the observations. To emphasize the differences between the $\Lambda$CDM model and other dark energy models, we set $\triangle$AIC and $\triangle$BIC for the $\Lambda$CDM model to zero, with the corresponding values presented in Table \ref{tab4}.
The AIC is defined as:

\begin{equation}
AIC=\chi^{2}_{min}+2d
\end{equation}
where $d$ represents the model's number of free parameters. The BIC is defined similarly by

\begin{equation}
BIC=\chi^{2}_{min}+ d Ln N
\end{equation}

The value of \( N \) refers to the number of data points used in the analysis. In general, a lower AIC/BIC value suggests better alignment with the data. To determine which model is preferred based on the given measurements, one usually computes the differences \(\Delta AIC = AIC_{2} - AIC_{1}\) and \(\Delta BIC = BIC_{2} - BIC_{1}\).

 In this context, larger values of \( |\Delta \mathrm{BIC}| \) or \( |\Delta \mathrm{AIC}| \) provide stronger evidence against the model \cite{ref86}. The outcome is considered \emph{positive} when the absolute difference in AIC or BIC, denoted as \( |\Delta \mathrm{AIC}| \) or \( |\Delta \mathrm{BIC}| \), satisfies the following criteria: \( |\Delta \mathrm{AIC}| \leq 2 \) denotes significant observational support, \( 4 \leq |\Delta \mathrm{AIC}| \leq 7 \) indicates moderate support, \( |\Delta \mathrm{AIC}| \geq 10 \) reflects a lack of observational support. For the BIC, the evidence is considered \emph{positive} when \( 2 \leq |\Delta \mathrm{BIC}| \leq 6 \), \emph{strong} when \( 6 < |\Delta \mathrm{BIC}| \leq 10 \), and \emph{very strong} when \( |\Delta \mathrm{BIC}| > 10 \).\\
 
Table \ref{tab4} presents the AIC and BIC values for the $w$CDM and $\Lambda$CDM models across various combined datasets. The values of $\Delta \text{AIC} = \text{AIC}_{w\text{CDM}} - \text{AIC}_{\Lambda \text{CDM}}$ and $\Delta \text{BIC} = \text{BIC}_{w\text{CDM}} - \text{BIC}_{\Lambda \text{CDM}}$ are all positive, indicating that the $\Lambda$CDM model is statistically preferred over the $w$CDM model in our analysis. From DESI BAO+BBN and DESI BAO+BBN+OHD datasets, the $\Delta \text{AIC}$ value  is close to 2, indicating that $w$CDM model favored observational support in the comparison $\Lambda$CDM. Meanwhile other two given dataset, our model shows moderated consistent. In contrast, the $w$CDM model is favored for the corresponding range $2 \leq |\Delta \mathrm{BIC}| \leq 6$, particularly when using the datasets DESI BAO+BBN and DESI BAO+BBN+OHD. Based on these statistical analyses, our dynamical dark energy model remains consistent with the standard model of cosmology. Moreover, our findings are consistent with prior studies \cite{ref87aa,ref87a,ref87b,ref87c}, which also report slightly higher BIC and smaller AIC values for dynamical dark energy models due to the inclusion of additional parameters. Therefore, while $\Lambda$CDM provides a better fit in terms of information criteria, the $w$CDM model remains compatible with the data and offers valuable insights, especially in the context of addressing current observational tensions.

\begin{table*}[hbt!]
     \caption{The difference, $\Delta \text{AIC} = \text{AIC}_{w\text{CDM}} - \text{AIC}_{\Lambda \text{CDM}}$ and $\Delta \text{BIC} = \text{BIC}_{w\text{CDM}} - \text{BIC}_{\Lambda \text{CDM}}$ for $w$CDM model with respect to $\Lambda$CDM  from all considered data sets.}
     \label{tab4}
     \scalebox{0.85}{
 \begin{tabular}{lcccc}
  	\hline
    \toprule
   \textbf{Dataset }&\;\textbf{DESI BAO+BBN}\;&\; \textbf{DESI BAO+BBN+OHD}\;& \;\;\;\;\;\textbf{DESI BAO+BBN+OHD+SN$^+$}\;\; & \textbf{DESI BAO+BBN+OHD+SN$^+$\&SHOES}     
   \\ \hline
      \textbf{Model} & \textbf{$\bm{w}$CDM}\,&\textbf{$\bm{w}$CDM}\,&\textbf{$\bm{w}$CDM}\,&\textbf{$\bm{w}$CDM}\vspace{0.1cm}\\
&\textcolor{teal}{\textbf{$\bm{\Lambda}$CDM}}\, & \textcolor{teal}{\textbf{$\bm{\Lambda}$CDM}}\, & \textcolor{teal}{\textbf{$\bm{\Lambda}$CDM}}\, & \textcolor{teal}{\textbf{$\bm{\Lambda}$CDM}} 
          \\ \hline

\vspace{0.1cm}
{\boldmath$\rm AIC$}&$ 21.76$ &$ 36.16$ &$1464.44 $ &$ 1359.38$\\

&\textcolor{teal}{$19.56$} &\textcolor{teal}{$34.10$} &\textcolor{teal}{$1460.72$} &\textcolor{teal}{$1354.36$}\\
\hline
\vspace{0.1cm}
{{\boldmath$\rm \Delta AIC$}}&$ 2.20$ &$ 2.06$ &$ 4.44  $& $ 5.02$\\

\hline
\vspace{0.1cm}
{{\boldmath$\rm BIC$}}&$ 24.31 $&$43.56$&$1484.43 $ &$1381.43$ \\
  
& \textcolor{teal}{$ 21.47$}& \textcolor{teal}{$ 39.65$} &\textcolor{teal}{$ 1475.71$}&\textcolor{teal}{$1370.75 $}\\
\hline
\vspace{0.1cm}
{{\boldmath$\rm \Delta BIC$}}&$2.84 $& $3.91 $&$8.72 $ &$ 10.68$\\

 \hline
 \hline
\end{tabular}
}
\end{table*}

\section{Conclusions}

In conclusion, this study provides a detailed analysis of the cosmological models addressing the Hubble tension, with a particular focus on the comparison between the $\Lambda$CDM and $w$CDM models. By incorporating a variety of datasets, including the recent DESI BAO, BBN, OHD and SN$^{+}$\&SH0ES data, we have explored the impact of these observations on the dark-energy EoS and the Hubble constant. Our results show that the $w$CDM model offers a promising resolution to the Hubble tension, reducing the statistical discrepancy in the $H_0$ values obtained from different methods. Specifically, the $w$CDM model consistently yields a higher value for $H_0$ than the $\Lambda$CDM model, with the inclusion of more data sets further mitigating the tension.\\

In this paper, we have constrained a baseline and several derived parameters for the $w$CDM model, an extension of the $\Lambda$CDM model, using various combinations of datasets, including DESI BAO, BBN, OHD, and SN$^+$\&SH0ES. Our analysis shows that incorporating SH0ES data with DESI BAO+BBN+OHD+SN$^+$ leads to a transition of the universe from the phantom to the quintessence regime foam of dark energy, also, we find a $1.6\sigma$ deviation of $w_{de0}$ from the cosmological constant with DESI BAO+BBN+OHD+SN$^+$ dataset. However, other considered dataset combinations do not show any significant deviation of $w_{de0}$ from the cosmological constant. Besides, we have observed that the $w$CDM model estimates the higher values of the Hubble constant  $H_0 =  70.50\pm 3.70 (71.56 \pm 0.79 ) \rm{km} \rm{s}^{-1} \rm{Mpc}^{-1}$ from DESI BAO+BBN(DESI BAO+BBN+OHD+SN$^+$\&SH0ES ) data analysis, respectively. Furthermore, the $w$CDM model helps alleviate the Hubble tension by approximately $2.1\sigma (0.8\sigma)$ from  DESI BAO+BBN+OHD (DESI BAO+BBN+OHD+SN$^+$\&SH0ES). The derived $H_0$ values both consider data combinations without SN$^+$ and are consistent with SH0ES measurements. Overall, we conclude that combining DESI BAO data with BBN, OHD, and PP\&SH0ES datasets effectively relaxing the existing  tension both Hubble constant ($H_0$) and the absolute magnitude ($M_B$)  within the $w$CDM model. Moreover, our results show strong consistency with those results by the DESI collaboration for the $w$CDM model, especially when observational Hubble parameter data (OHD) are taken into account. Furthermore, our analysis, using both the Akaike Information Criterion (AIC) and the Bayesian Information Criterion (BIC), indicates that the $w$CDM model provides a better fit to the observational data compared to the standard $\Lambda$CDM model. This suggests that a dynamical dark energy model may be more in line with current observations, especially when considering the latest data from DESI and other large-scale surveys.\\ 

The $\omega$CDM model is not always derived from a covariant conservation law of a specific physical field, particularly when $\omega \ne -1$.  $\omega$CDM serves as a phenomenological model for testing deviations from $\Lambda$CDM.  It enables cosmologists to ask, "What if dark energy isn't a cosmological constant?"  Observations (e.g., CMB, supernovae, BAO) help to constrain $\omega$.  Currently, they are compatible with $\omega \approx -1$, therefore $\Lambda$CDM suits best. Using $\omega$CDM is not deceptive as long as we recognize its limits.  $\omega$CDM is not a comprehensive physical model, yet it is not deceptive when used as a phenomenological tool. 
Our study on the considered model (1) is important for several compelling scientific reasons. Here is a detailed explanation of its significance:\\
(a) Testing Dark Energy Beyond $\Lambda$CDM: The $\omega$CDM model extends the standard $\Lambda$CDM cosmology by allowing the dark energy equation of state parameter $\omega$ to deviate from $-1$.  If $\omega \ne -1$, it indicates dynamic dark energy, implying physics beyond the cosmological constant.  Our work evaluates the $\omega$CDM model using current, high-precision data.\\
(b) Independent Validation Without CMB: By focusing on CMB-independent datasets, such as the latest DESI BAO measurements, our work avoids relying on early-universe physics alone. This is crucial for cross-validating our understanding of cosmic acceleration from low-redshift (late-time) observations. CMB measurements (e.g., from Planck) are incredibly powerful, but they probe the universe at $z \sim 1100$. Our study adds complementary constraints from the much later universe ($z < 3$ ), where dark energy dominates.\\
(c) DESI BAO: A Game-Changer in Precision Cosmology: The DESI experiment considerably increased the precision of BAO measurements.  BAO functions as a standard ruler, enabling exact measurement of the expansion rate $H(z)$ and distance scales $D_{M}(z)$, $D_{A}(z)$.  Our study utilizes DESI's extensive redshift sampling and great accuracy to set strict limitations on $\omega$.  This helps to determine if $\omega = -1$.  Look for evidence of redshift evolution or divergence from the constant $\omega$. \\
(d) Constraining Cosmological Tensions: Our analysis may also shed light on current tensions in cosmology, such as the Hubble tension, i.e, differences in $H_{0}$ between local and CMB-inferred measurements. If our CMB-independent constraints support a different value of $\omega$ or $H_{0}$, that could point to new physics (e.g., evolving dark energy, early dark energy, modified gravity) or point to systematic issues in some datasets
 \\

The authors of the study \cite{ref88} released BAO data from more than 14 million galaxies and quasars from the DESI Data Release 2 (DR2), which were obtained over three years of operation.  The results are well explained by a flat $\Lambda$CDM model, however the parameters selected by BAO are in mild, 2.3$\sigma$ tension with those derived from the CMB. The DESI results are consistent with the acoustic angular scale $\theta_{*}$ that is well-measured by Planck. In the study \cite{ref89}, the authors presented the DESI 2024 galaxy and quasar BAO measurements using over 5.7 million unique galaxy and quasar redshifts in the range $0.1 < z <2.1$. They presented a re-analysis of SDSS BOSS and eBOSS results applying the improved DESI methodology and find scatter consistent with the level of quoted SDSS theoretical systematic uncertainties.\\ 
Furthermore, in our study, we investigate the impact of various combinations of CMB-independent datasets, including the latest DESI BAO measurements, on the equation of state (EoS) of dark energy and other cosmological parameters using the dynamical dark energy model ($w$CDM).   We use observational datasets including DESI BAO, BBN, Observational Hubble Data (OHD), and Pantheon Plus (SN$^+$) $\&$ SH0ES to restrict the model's free parameters, assuming a constant EoS parameter for dark energy.  Our analysis explores the variations of the $w$CDM model from the typical $\Lambda$CDM scenario and assesses their implications for cosmic tensions, especially the $H_0$ tension. The $w$CDM model is a simple expansion of the standard $\Lambda$CDM model, where the dark energy equation of state parameter $w$ can vary from the cosmological constant value ($w = -1$).  In the $w$ CDM model:  (i) Dark energy pressure: $p_{DE} = \omega \rho_{DE}$. (ii) $\omega = -1$ recovers $\Lambda$CDM. (iii) Dark energy evolves differently over time for $\omega \neq -1$.  \\

In the case of the $w$EDE model, the tension in the matter density estimation reported in previous studies disappears when incorporating BAO measurements from DESI while excluding the full CMB datasets. In summary, the BAO data obtained by DESI and DES exhibits significant potential in constraining cosmological parameters, especially when combined with independent estimates of the sound horizon and Hubble constant.

\section*{Declaration of competing interest}
	\noindent 
We wish to confirm that there are no known conflicts of interest
associated with this publication and there has been no significant financial
support for this work that could have influenced its outcome.

\section*{Data availability}
	\noindent 
No data was used for the research described in the article.

\section*{acknowledgments}
\noindent  
 The authors (A. Dixit \& A. Pradhan) are thankful to IUCAA, Pune, India for providing support and facility under Visiting Associateship program.  M. Yadav is sponsored by a Junior Research Fellowship from the University Grants Commission, Government of India (CSIR/UGC Ref.\ No.\ 180010603050). The authors are appreciative to the Reviewer and Editor for their informative remarks, which improved the manuscript in its current form.

\begin{center}
   \textbf{Appendix I : Triangle Countor  } 
\end{center}

In this appendix, we present the one- and two-dimensional marginalized distributions (at 68\% and 95\% CL) for the parameters   $\omega_b, \omega_{cdm}, H_0, w_{de0}, M_B $ and $\Omega_m $ within the $\Lambda$CDM model. We observe a slight impact of the DESI BAO data on the estimation of these cosmological parameters.

\begin{figure*}[hbt!]
    \centering
    \includegraphics[width=0.7\linewidth]{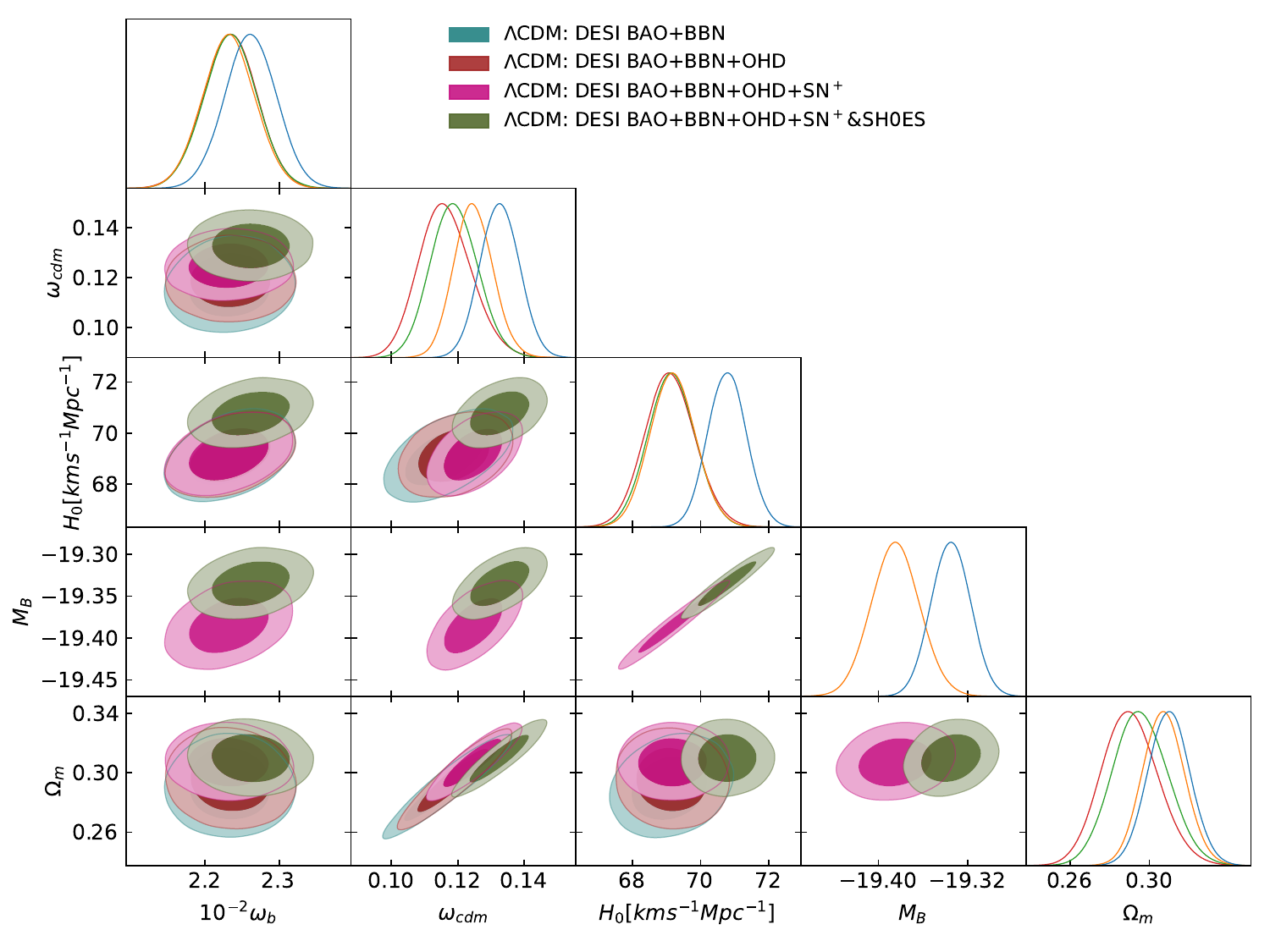}
    \caption{One-dimensional and two-dimensional marginalized contour plots at 68 \%and 95\% C.L for various parameters obtained from
 DESI BAO+BBN, DESI BAO+BBN+OHD, DESI BAO+BBN+OHD+SN$^+$, and DESI BAO+BBN+CC+SN$^+$\&SH0ES datasets with the $\Lambda$CDM model.}
    \label{fig5}
\end{figure*}

\end{document}